\documentclass[manuscript]{aastex}
\usepackage{spr-astr-addons}
\usepackage{epstopdf}

\RequirePackage{color}

\begin{document}
\textit{Accepted for publication in JPSA, Vol.7, no. 4, 2017}\\

\title{On the mystery of the interpulse shift in the Crab pulsar}
%% Running heads
\shorttitle{Shift of Interpulse in Crab Pulsar} \shortauthors{V.M.~Kontorovich and S.V.~Trofymenko}

\author{V.M.~Kontorovich \altaffilmark{1,2}}
\and \author{S.V.~Trofymenko \altaffilmark{1,3}}

\altaffiltext{1}{Karazin Kharkov National University, 4 Svobody
	Sq., Kharkov 61022, Ukraine,} 

\altaffiltext{2}{Institute of Radio Astronomy of National Academy
	of Sciences of Ukraine, 4 Art Str., Kharkov 61002,}

\altaffiltext{3}{Akhiezer Institute for Theoretical Physics of NSC KIPT, 1 Akademicheskaya Str., Kharkov 61108, Ukraine.}

\begin{abstract}
A new mechanism of radiation emission in the polar gap of a pulsar is proposed. It is the curvature radiation which is emitted by returning positrons moving toward the surface of the neutron star along field lines of the inclined magnetic field and reflects from the surface. Such radiation interferes with transition radiation emitted from the neutron star when positrons hit the surface. It is shown that the proposed mechanism may be applicable for explanation of the mystery of the interpulse shift in the Crab pulsar at high frequencies discovered by Moffett and Hankins twenty years ago.
\end{abstract}

%% Keywords
\keywords{neutron stars, pulsar PSR B0531+21, non-thermal radiation mechanisms, shift of interpulse, coherent emission, radiation energy spectrum}

\section{Introduction}\label{s:intro}

It is commonly accepted that a pulsar is a neutron star with the radius of about ten kilometers which creates intense magnetic field around itself. In the vicinity of the surface of the star such field is often supposed to resemble the field of a magnetic dipole. The star quickly rotates around the axis which does not coincide with the direction of its magnetic moment. Such rotation causes the division of the space around the star into two regions (see e.g. \cite{b27,Manchester}). The first one is situated within the cylinder (light-cylinder) of radius $R_c=c/\Omega$ (where $c$ is the speed of light and $\Omega$ is the pulsar rotation angular frequency) with the axis coinciding with the pulsar rotation one. The second region lies outside this cylinder. It leads to the division of the magnetosphere into the regions of closed and opened magnetic field lines. The first ones are situated completely within the light-cylinder in the region filled with co-rotating plasma of electron-positron pairs. The second ones emerge from the surface of the star in the vicinity of the magnetic poles and intersect the light-cylinder.      

Quick rotation of the highly magnetized star leads to generation of intense electric field around it. In the vicinity of the magnetic poles this field has a longitudinal component parallel to the magnetic field. It accelerates the electrons from the surface of the star along the curved opened magnetic field lines up to the value of the Lorentz-factor of the order of $10^7$. The hard part of curvature radiation emitted by such electrons generates the cascade of electron-positron pairs, which form the magnetospheric plasma (see e.g. \cite{b3}). During the acceleration process the electrons move inside a polar gap \citep{b45,b41,b2}, between the surface of the star and the pair plasma. 

We accept that the polar gaps are also the sources of electromagnetic radiation which has relatively narrow angular distribution around the magnetic axis but occupies a wide range of frequencies: from radio to gamma. Its origin is attributed to radiation produced by electrons accelerated in the polar gap by the electric field and penetrating into the magnetospheric plasma. It seems that a whole lot of mechanisms of radiation production by such electrons take place in this case replacing each other in different frequency ranges. Such mechanisms have been studied in a large number of papers (see, e.g. books and reviews \cite{b27,Manchester,b18,b2,b8,b3,Kaspi,Eilek2016} and references therein).    

Twenty years ago a thorough study of the average shapes of radiation pulses (average light curves) arriving from the pulsar in Crab nebula was made in \citep{b22}. During such investigation the observation data was collected in wide range of frequencies: from hundreds MHz to a hundred keV. Recently these results were confirmed and supplemented \citep{b7}. Here we reproduce (with some additional marks) a part of the figure from \citep{b22} which represents the results of measurements of the phase dependence of the average registered radiation intensity during one period of the pulsar rotation (fig.1). The average light curves are presented for different radiation frequencies.

\begin{figure}[h]
	\begin{center}
	\includegraphics[width = 80mm]{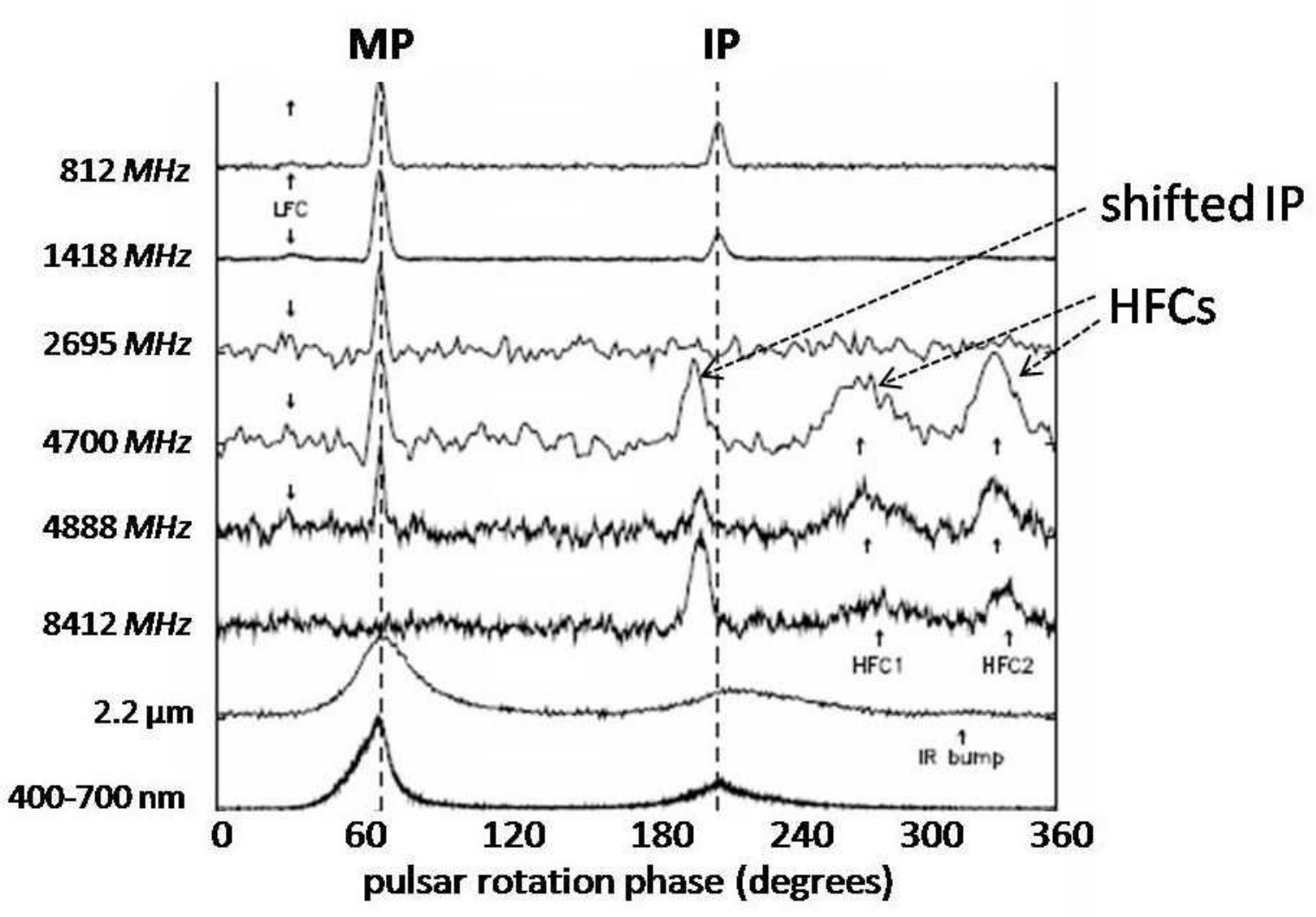}
	\end{center}
	\caption{Average light curves obtained at multi-frequency observations in \citep{b22}. The shifted (to about $7^{\circ}$) interpulse and high-frequency components are marked. With gratitude to the authors}
	\label{Fig1}
\end{figure}

At frequencies lower than 1.5 GHz two distinct pulses, registered during one period of rotation of the star, are seen on fig.1. Such pulses, the main pulse (MP) and interpulse (IP), are believed to originate from different magnetic poles of the Crab pulsar. Here we will assume that the magnetic axis of the star is nearly orthogonal to its rotation axis.  

At frequencies around 3 GHz IP disappears. It appears again at higher frequencies having phase sift $\delta$ of about $7^{\circ}$ comparing to the initial IP at lower frequencies. Moreover, two more distinct pulses, known as high-frequency components (HFCs), appear at the same frequencies. At some higher frequency HFCs disappear and IP restores its previous position. Let us note that MP disappears as well at the frequency of about 8 GHz appearing again at higher frequencies without any phase shift \citep{b22}.

Since the discovery of these peculiar features of radio emission by the Crab pulsar no theoretical explanation of them has been presented. In the present work we consider a mechanism of radio emission which may be responsible for the shift of IP in the Crab pulsar. At the heart of this mechanism lies the reflected from the pulsar surface curvature radiation by returning positrons\footnote{In the regions with the opposite direction of the accelerating electric field in the gap the same may refer to electrons moving to the star surface} moving in the polar gap toward the surface of the star. We show that such mechanism provides the possibility of explaining the IP phase shift. The version of the origin of HFCs is discussed in \citep{b31} also involving the reflection from the star surface and applying the idea of nonlinear Raman scattering of the positron radiation on the excitations of the pulsar surface. The idea of reflection from the pulsar surface is applied here for consideration of HFCs since they appear at the same frequencies as the IP shift, which presently does not have other explanation. For independent treatment of the HFCs, not connected with the IP shift, see \citep{Petrova}.

%\begin{figure}
%	\includegraphics[width = 70mm]{Fig1.eps}
%	\caption{Schematic picture of electron and positron radiating at the same frequency (the curvature radiation spectrum maximum) corresponding to the same Lorentz-factor}
%	\label{Fig1}
%\end{figure}

\section{Radiation by returning positrons as the source of the shifted IP}

The explanation of MP disappearance was proposed in \citep{b14}. Such proposal was made on the basis of consideration of non-relativistic radiation mechanism during longitudinal acceleration of electrons in the polar gap as the mechanism of low-frequency radiation emission. When the accelerated electron reaches relativistic velocities this mechanism weakens and turns off at rather high frequency. As was shown there, the value of such frequency is proportional to $\sqrt{B_{||}}$, where $B_{||}$ is the  magnetic field component parallel to the rotation axis in the vicinity of the magnetic pole of the star. The disappearance of IP may be attributed to the same physical reasons. The lower value of the frequency at which it occurs, comparing to the one in the case of MP, can be caused by the lower value of $B_{||}$ on the magnetic pole responsible for IP production (for details see \cite{aasp}).  

Anyway, the radiation mechanism responsible for production of the shifted IP at high frequencies should differ from the one dominating in the case of unshifted IP at lower frequencies (such idea correlates with the one expressed in \cite{b7}).

As the origin of the shifted high-frequency IP we consider the radiation by positrons moving toward the surface of the pulsar in the polar gap. Such positrons can be returned from the lower layers of magnetospheric plasma (just after the corresponding $e^+e^-$ pairs are created) by the same electric field which accelerates the electrons outward the star. The detailed consideration of such return motion of positrons in a pulsar magnetosphere is presented in \citep{b47} and references therein. 

We assume that in the considered range of frequencies of the order of several GHz the surface of the star acts as ideal conductor reflecting all the incident radiation. Moving along curved magnetic field lines towards the surface of the star the positrons emit curvature radiation (CR), just like the electrons moving in the opposite direction \citep{b10}. Since the positrons are ultra relativistic, such radiation is emitted predominantly in the direction close to their velocities. At sufficiently high frequencies it reflects from the surface and propagates outward the star. Moreover, when the positrons hit the surface the so-called transition radiation \footnote{Transition radiation is the radiation which is generated when a moving charged particle traverses the border between media with different dielectric properties. It has wide and nearly constant spectrum covering the range from the long-wavelength part of the radio band and up to X-ray band for ultra relativistic particles. For details see \citep{b32,b28}} is generated in the direction outwards the surface as well. The total radiation by positrons is the result of interference of such transition radiation with the reflected curvature radiation\footnote{According to \citep{Dyks2006} and \citep{Wright}, the reflection from the neutron star surface was discussed in the report \citep{Michel} on the conference in Zielona Gora, presumably, in connection with the problem of the drifting subpulses structure. Let us also remark that an interesting question concerning the so-called reversible radio emission (see i.e. \cite{Melikidze}) does not have direct analogy with our problem}.    

In order to obtain the angular shift between the predominant direction of the direct radiation by electrons, which is believed to be the origin of the unshifted low-frequency IP, and the corresponding direction of the reflected radiation by positrons (shifted IP) one more essential assumption is needed. It is the requirement that the magnetic axis on the surface of the star should be inclined at some angle $\delta/2$ (of about several degrees) to the surface normal (fig.2).
\begin{figure}[h]
	\begin{center}
		\includegraphics[width = 50mm]{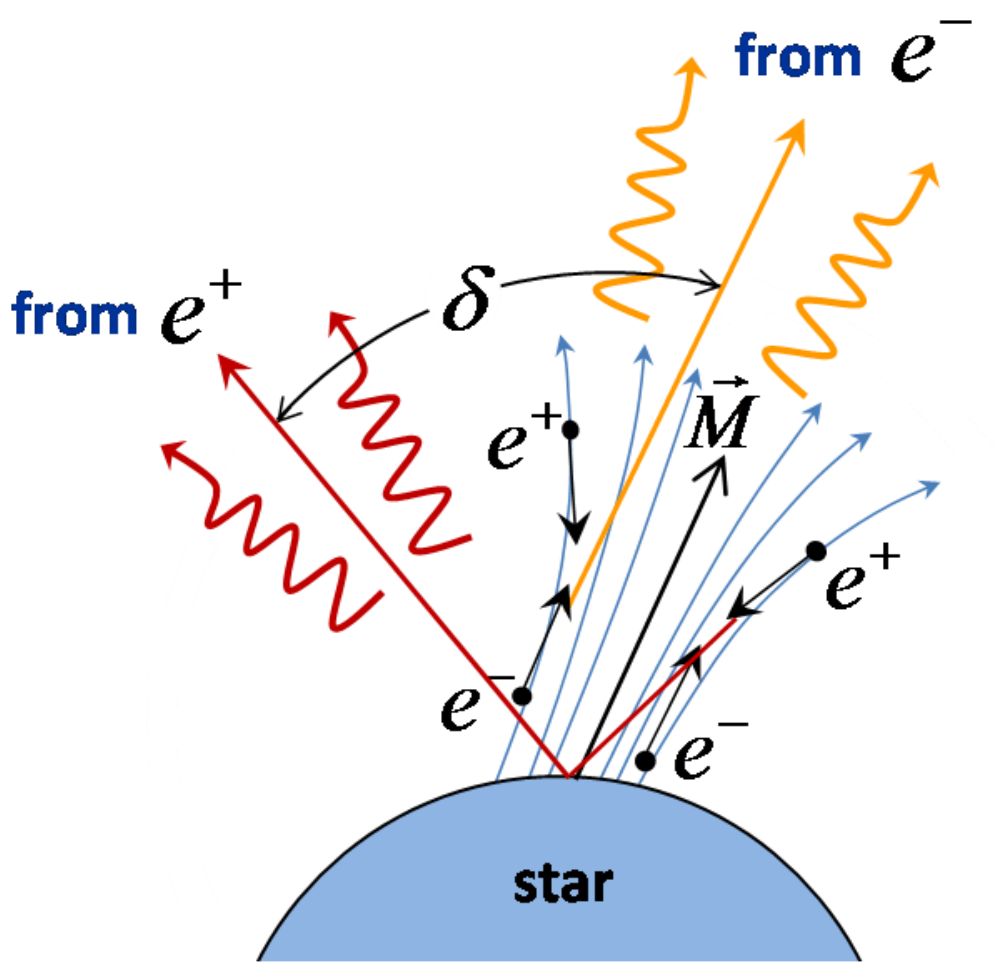}
	\end{center}
	\caption{Schematic picture of motion and radiation by electrons and positrons in the polar gap of the pulsar in the case of a tilted magnetic axis. The directions of radiation by electrons and reflected radiation by positrons are shifted at the angle $\delta$ of mirror reflection}

	\label{Fig2}
\end{figure}

The direction of the magnetic axis defines the average direction of radiation by electrons in the polar gap. In the considered case the predominant angular domain of concentration of radiation by positrons is tilted to the angle of about $\delta$ from the magnetic axis. It happens due to the fact that both the average directions of transition radiation\footnote{In the case of a relativistic charged particle oblique incidence upon a conducting surface the major part of backward TR pulse is concentrated close to the direction of mirror reflection with respect to the direction of the particle velocity} and reflected CR in this case are the directions of mirror reflection of the magnetic axis with respect to the surface of the star.  

\section{The model of a returning positron motion}

In order to investigate the properties of radiation produced by the positrons moving toward the surface of the star and hitting it, it is necessary to adopt a certain (naturally, simplified) picture of their motion. Following the ideas elaborated in \citep{b33,b10} for radiation by electrons, for certainty, we will mainly consider radiation by positrons moving along outer open magnetic field lines. Such lines are situated in the vicinity of the boundary between the region of open and closed lines and have significant curvature. However, they are supposed not to belong to the immediate vicinity of this boundary and still have significant accelerating electric field component along them. CR produced by such positrons is more intensive than the one emitted by the particles moving along more straight (with larger curvature radius) field lines situated closer to the magnetic axis (see discussion in sec. 6 involving consideration of radiation coherence). Nevertheless, for estimation of the total positron radiation flux in sec. 5 and 6 we will take into account the contribution from the positrons on all the opened field lines. In this section the important effect of radiation coherence will be included into our consideration as well. 

Firstly, let us note that CR emitted by a positron can reflect from the surface of the star and propagate to the observer only if the emission occurs within some effective part of the positron trajectory (further we will call it effective path). On fig.3 this part leans on the angle $\alpha_0$. The value of the positron Lorenz-factor $\gamma$ within the effective path is restricted by its minimum $\gamma_{min}$ and maximum $\gamma_{max}$ values. The first one depends on the frequency and the curvature radius of the positron trajectory (see formula (\ref{eq6}) with the substitution $\gamma\rightarrow\gamma_{min}$). The second one is defined by the relative position of the observation direction and the magnetic axis (for details see sec. 5). Since CR emission by a relativistic positron occurs in the direction close to its velocity, CR emitted prior to this region does not reflect from the surface and, hence, is not observed. By $\alpha$ we denote here the angular coordinate of the positron which is counted from the point of the beginning of its motion along the effective path. 

For the sake of simplicity of calculations we assume that the positron moves along an arc of a circle of radius $R\sim 10^7 cm$. In this case effective path corresponds to the angle $\alpha_0 \simeq \arccos(1-R_*/R)-R_*/R$, where $R_*\sim 10^6 cm$ is the radius of the pulsar, and has the value of about $3.5R_*$. In fact, it is just the highest estimation for the effective path since the radiation emitted in the direction close to the `limit' one (see fig.3) may not escape from the pulsar at all. In order to reach the observer the reflected radiation should propagate in the direction closer to the magnetic axis, which requires the emission to occur at lower altitudes (at larger $\alpha$).        

\begin{figure}[h]
	\begin{center}
		\includegraphics[width = 80mm]{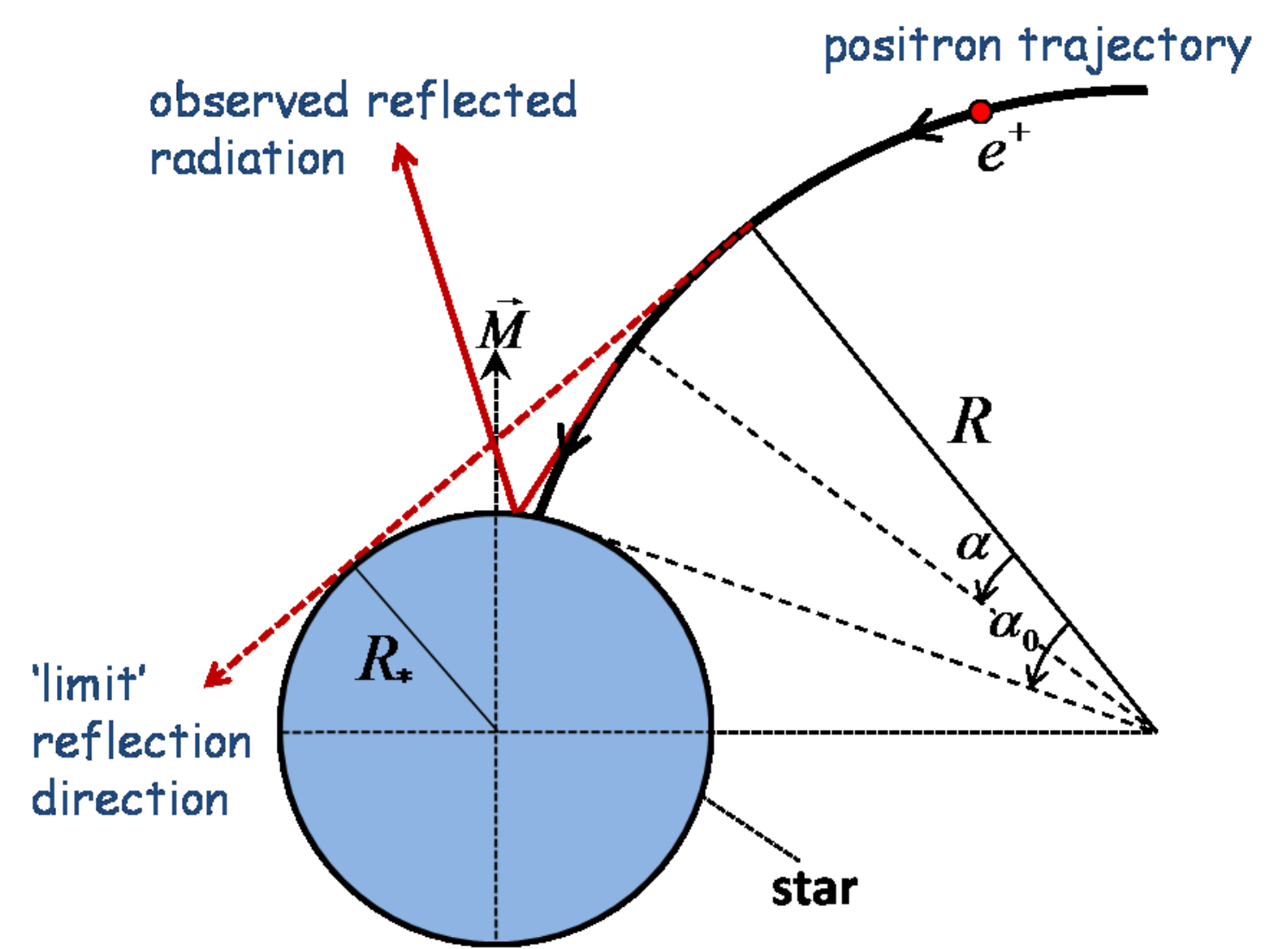}
	\end{center}
	\caption{Schematic picture of a return positron motion and radiation along the curved magnetic field line.  $\alpha_0$ is the angle corresponding to the effective path, $\alpha$ is the current positron angular coordinate. The solid red line shows the direction of the reflected radiation emitted by the positron at its current position, while the dashed one shows the direction of radiation (emitted at $\alpha=0$) which just begins being reflected from the surface \citep{aasp}}
	\label{Fig3}
\end{figure}

In the region corresponding to the effective path we use the following simplified model of accelerating field:
\begin{equation}\label{eq1}
E(\alpha)=E_1 (\alpha/\alpha_1-\alpha) \theta(\alpha_1-\alpha)+
\end{equation} 
$$
\bigl ( E_1+E_2 ( \alpha-\alpha_1 ) \bigr )
(1-\alpha) \theta(\alpha-\alpha_1),
$$ 
where $\theta(x)$  is the step function, which is equal to zero for $x<0$  and to unit for $x>0$.

In this model the field firstly slowly and linearly increases in the first part of the effective path ($\alpha<\alpha_1$), which lies in the vicinity of the magnetospheric plasma, and then changes parabolically reaching some maximum value and turning to zero on the surface of the star ($\alpha=1$). 
In (\ref{eq1}) $\alpha$ and $\alpha_1$ are measured in the units of  $\alpha_0$. 
The linear section corresponds to smooth penetration of the field into the lower magnetosphere, cf. \citep{b24}, the parabolic -- to a simplified version of known models, see \citep{b2, b8}. $E_1$ and $E_2$ are some constants defining the magnitude of the electric field. 

The equation defining the dependence of the positron Lorenz-factor $\gamma$ on $\alpha$ is 
\begin{equation}\label{eqgamma}
d\gamma/d\alpha=eE(\alpha)R/mc^2,
\end{equation}
 which can be derived from the well-known equation for a particle energy gain in external electric field: 
\[
d\varepsilon/dt=e\bf{Ev}.
\]
 The solution of this equation with the field given by (1) leads to the following expression for the Lorenz-factor:
\begin{equation}\label{eq3}
\gamma(\alpha)=\theta{(\alpha_1-\alpha)}{\{\gamma(0)+f_1(1-\alpha_1)\alpha^2/ 2\alpha_1^2 \}}+
\end{equation}
$$
+\theta{(\alpha-\alpha_1)}{\{\gamma(0)+f_1(1-\alpha_1)/2+\delta \gamma(\alpha)}\},
$$
$$
\delta \gamma = (f_1/\alpha_1-f_2) (\alpha-\alpha_1) +
$$
$$
+ \bigl ( f_2 ( 1+\alpha_1 ) - f_1 \bigr ) \bigl ( \alpha^2 - \alpha^2_1 \bigr )/2 \alpha_1 - f_2(\alpha^3 - \alpha^3_1 )/3\alpha_1,
$$  
where $f_{1,2}=e\alpha_0E_{1,2}R\alpha_1/mc^2$.

In the next section it will be shown that it is possible to choose the values of $E_1$ and $E_2$ in such way that radiation emitted by a positron in the region of the linear growth of the field belongs to radio band. Moreover, the appearance of the shifted IP in the vicinity of a certain frequency (about 5 GHz, see fig.1) can be realized in the adopted model.  

\section{Spectral-angular characteristics of radiation by a returning positron}

In order to calculate the spectral-angular density of radiation emitted by a positron moving along a curved magnetic field line in the polar gap and falling on the surface of the star we use the method of images. Such method is applicable even in ultra relativistic case if the surface of the star  is approximately considered as flat. In this method the radiation emitted by the positron is considered as radiation emitted by its image moving inside the star the mirror symmetrically to the positron with respect to the surface (fig.4). When the positron enters the surface of the star it disappears from the point of view of the external observer. It happens due to the fact that the positron's charge in this case is completely screened by polarization charges and currents inside the star surface (which we consider here as a perfect conductor). In the terms of the method of images such disappearance of the positron is described as abrupt stop of the particle and its image at the same point on the surface \citep{b4,b32}. The charges of the positron and its image screen each other in this case which is analogous to the particles disappearance. 
\begin{figure}[h]
	\begin{center}
		\includegraphics[width = 70mm]{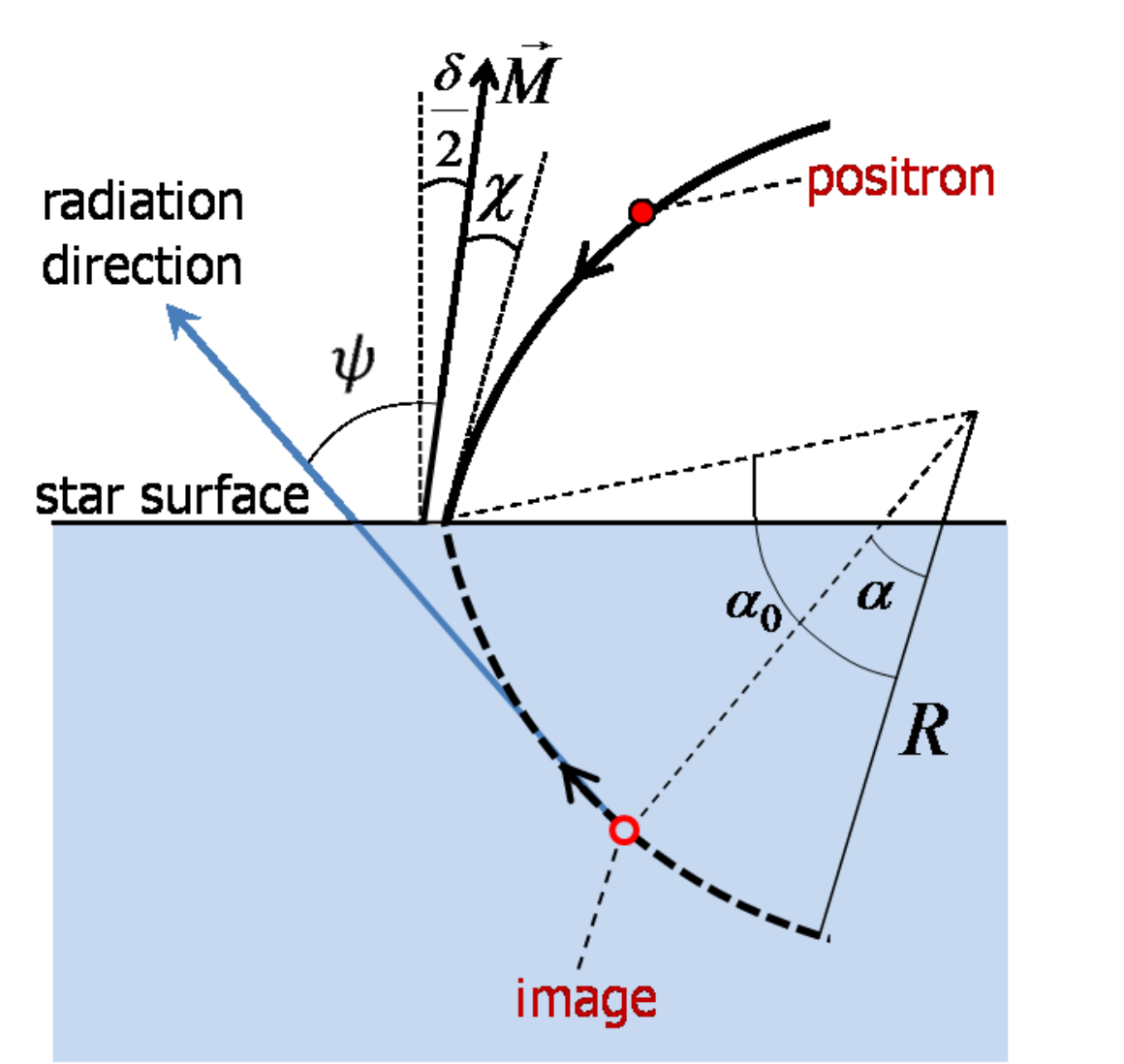}
	\end{center}
	\caption{The geometry of a positron and its image motion used for calculations. The radiation direction angle $\psi$  approximately corresponds to a certain value of the radiating particle Lorentz-factor and radiated frequency \citep{aasp}}
	\label{Fig4}
\end{figure}

In the considered method the radiation emitted by the image during its motion is analogous to the reflected curvature radiation of the positron. The bremsstrahlung generated by the image at its abrupt stop on the surface is analogous to transition radiation which occurs when the positron hits the surface. 

By $\chi$ we define the value of the angle between the line tangent to the magnetic field line on the surface of the star and the direction of the magnetic axis. The angle $\psi$ is the angle between the radiation observation direction and the magnetic axis. 

The spectral-angular density of radiation emitted by the positron's image can be calculated with the use of the well-known expression (e.g. see \cite{b9,b16}) for distribution of the energy radiated by a particle with the known law of motion ${\bf r=r}_0(t)$\footnote{This quantity can be expressed in terms of the positron acceleration $\bf\dot{v}$ as well through taking the integral with respect to $t$ in the expression (\ref{eq4}) by parts, see, e.g. \citep{b9}}: 
\begin{equation}\label{eq4}
\frac
{d^2 W}
{d\omega do} =
\frac{e^2
	\omega^2}{4\pi^2 c^3}
\Biggl |
\int
\limits_
{-\infty}^{\infty}dt
[ \mathbf{n}, \mathbf v(t)]
\exp \left \{ i\omega (t-\frac{\mathbf{n}\mathbf{r_0}(t)}{c}) \right \} \Biggr |^2.
\end{equation}
Here ${\bf v}(t)$ is the image velocity and $\bf n$ is a unit vector in the direction of observation. 

We do not take into account the radiation emitted when the positron changes its direction of motion (after the pair creation) under the impact of the electric field and starts its motion towards the surface. This is due to the fact that at the beginning of such motion the positron Lorenz-factor is close to unit. The radiation emitted in this region at relatively high considered frequencies (several GHz) is less intense than the curvature radiation generated during further positron motion towards the surface. 

Thus we consider radiation emitted by the positron (or rather its image) which approaches the effective path having some moderately relativistic initial velocity v$_0$. Then it gains energy accelerating along the curved magnetic field line up to standard values of particles Lorenz-factors in this case $\gamma\sim 10^6-10^7$ and hits the surface.

The radiation is predominantly concentrated in the vicinity of the plane of the positron motion. It is this plane which we will further consider radiation characteristics in. For the vector product in (\ref{eq4}) we can write: 
\[
[ \mathbf{n}, \mathbf v(t)]={\bf e_\perp}{\text v}\cdot \sin{(\alpha-\alpha_0-\chi-\delta+\psi)},
\] 
where $\bf e_\perp$ is a unit vector perpendicular to the plane of the positron motion. The integration over time in (\ref{eq3}) can be substituted by the one over $\alpha$ with the use of the relation ${\text v}dt=Rd\alpha$. The dependence of $t$ on $\alpha$ during the positron motion within the effective path ($0<\alpha<\alpha_0$) can be expressed with the use of (\ref{eqgamma}) as follows:
\[
t(\alpha)=\frac{R}{c}\int\limits_{0}^{\alpha}\frac{d\alpha'}{\beta(\alpha')},
\]   
in which $\beta(\alpha')={\text v(\alpha')}/c=\sqrt{1-1/\gamma^2(\alpha')}$.

Thus we obtain the following expression for spectral-angular density of radiation by a single positron in the considered case:

\begin{equation}\label{eq5}
\frac
{d^2 W}
{d\omega do} =
\frac{e^2{\text w}^2}{4\pi^2 c}
\Biggl |
i\frac{\beta_0 \sin(\eta(\psi))\exp[ i{\text w} \sin(\eta(\psi))]}{{\text w}[1-\beta_0 \cos(\eta(\psi))]}+
\end{equation}
$$
\int
\limits_
{0}^{\alpha_0}d\alpha
\sin(\alpha-\eta(\psi))
\exp \left \{ i{\text w} \bigg (\frac{ct}{R}-\sin(\alpha-\eta(\psi))\bigg ) \right \} \Biggr |^2,
$$
where $\eta(\psi)=\alpha_0+\chi+\delta-\psi$, ${\text w}=\omega R/c$ and $\beta_0={\text v}_0/c$.

Fig.5 shows the frequency dependence of the radiated energy at different observation angles $\psi$ calculated with the use of expression (\ref{eq5}).
The figure shows that in the direction defined by a certain angle $\psi$ the considered radiation is emitted in the vicinity of a certain frequency $\omega(\psi)$. This frequency corresponds to the maximum of radiation spectral distribution in the considered direction. Due to narrow angular distribution of the emitted radiation around the direction of the positron instantaneous velocity, it is a small part of the positron trajectory which is responsible for radiation at certain angle $\psi$. On this small part of the effective path, which corresponds to a certain value $\alpha=\alpha_{\psi}$, the positron has a certain Lorenz-factor $\gamma(\alpha_{\psi})$. Thus the characteristic frequency $\omega(\psi)$ of radiation reflected from the surface in the considered direction can be estimated as 
\begin{equation}\label{eq6}
\omega(\psi)\sim c\gamma^3(\alpha_{\psi})/R
\end{equation}
as a frequency of curvature (or synchrotron) radiation spectrum maximum. From fig.4 it is possible to derive the relation between $\psi$ and $\alpha_{\psi}$ as $\psi=\alpha_0+\delta+\chi-\alpha_{\psi}$.

The radiation emitted by positron at the very beginning of the effective path also has its characteristic frequency $\omega(\alpha_0+\delta+\chi)$ (since $\alpha_\psi=0$ in this case). However, this wave is emitted tangentially to the surface of the star (see fig.3) and propagates at large angle to the magnetic axis. Such wave, even if it escapes from the magnetosphere, does not hit the telescope. The same situation takes place for the waves emitted at lower altitudes ($0<\alpha\ll\alpha_0$) at the beginning of the effective path. Such waves, though being reflected from the surface of the star, still have rather large angle between the magnetic axis and the wave vector direction. The waves emitted at even lower altitudes after reflection from the surface propagate in the direction close enough to the one of the magnetic axis to be able to escape the polar gap through relatively rarefied layer of plasma. 

\begin{figure}[h]
	\begin{center}
		\includegraphics[width = 70mm]{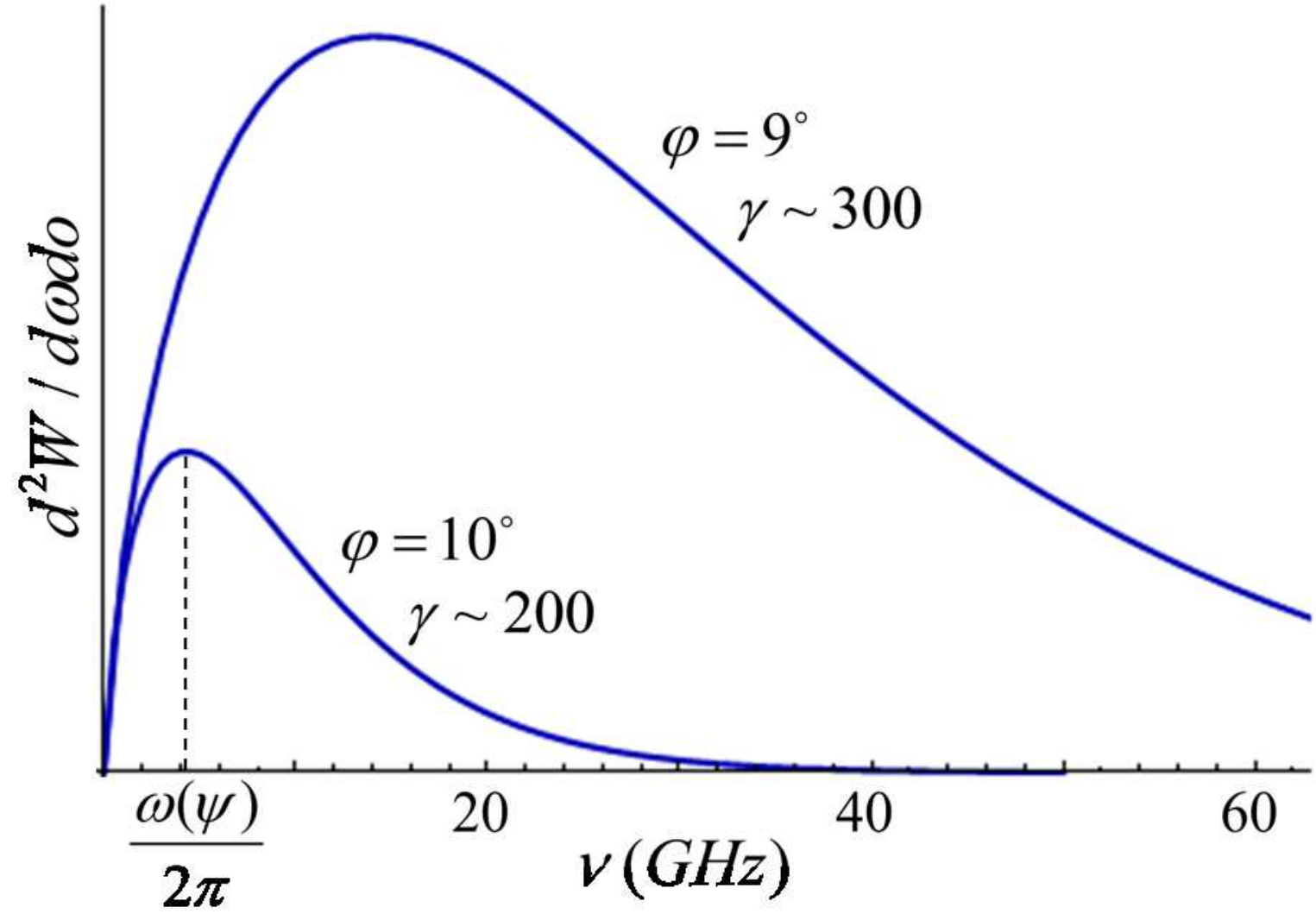}
	\end{center}
	\caption{Radiation spectral distribution (smoothed) for different values of $\varphi=\psi-\delta-2\chi$. The angle $\psi$ (see Fig.\ref{Fig4}) corresponds to the value (\ref{eq6}) of radiation frequency $\omega (\psi)$ and, hence, to a definite value of radiating positron Lorenz-factor $\gamma$. The values $f_1=4\cdot10^3$, $f_2=3\cdot10^7$ and $\alpha_1=1/2$ in the expression (\ref{eq3}) for $\gamma(\alpha)$ are used. The figure shows that at frequencies exceeding the one of the spectrum maximum, the radiation intensity rapidly decreases. The spectrum with the maximum at frequency $\omega_{min}$, at which the shifted IP appears, corresponds to the value $\gamma=\gamma_{min}$ (see (\ref{eq6_1})).}
	\label{Fig5}
\end{figure}

\begin{figure}[h]
	\begin{center}
		\includegraphics[width = 70mm]{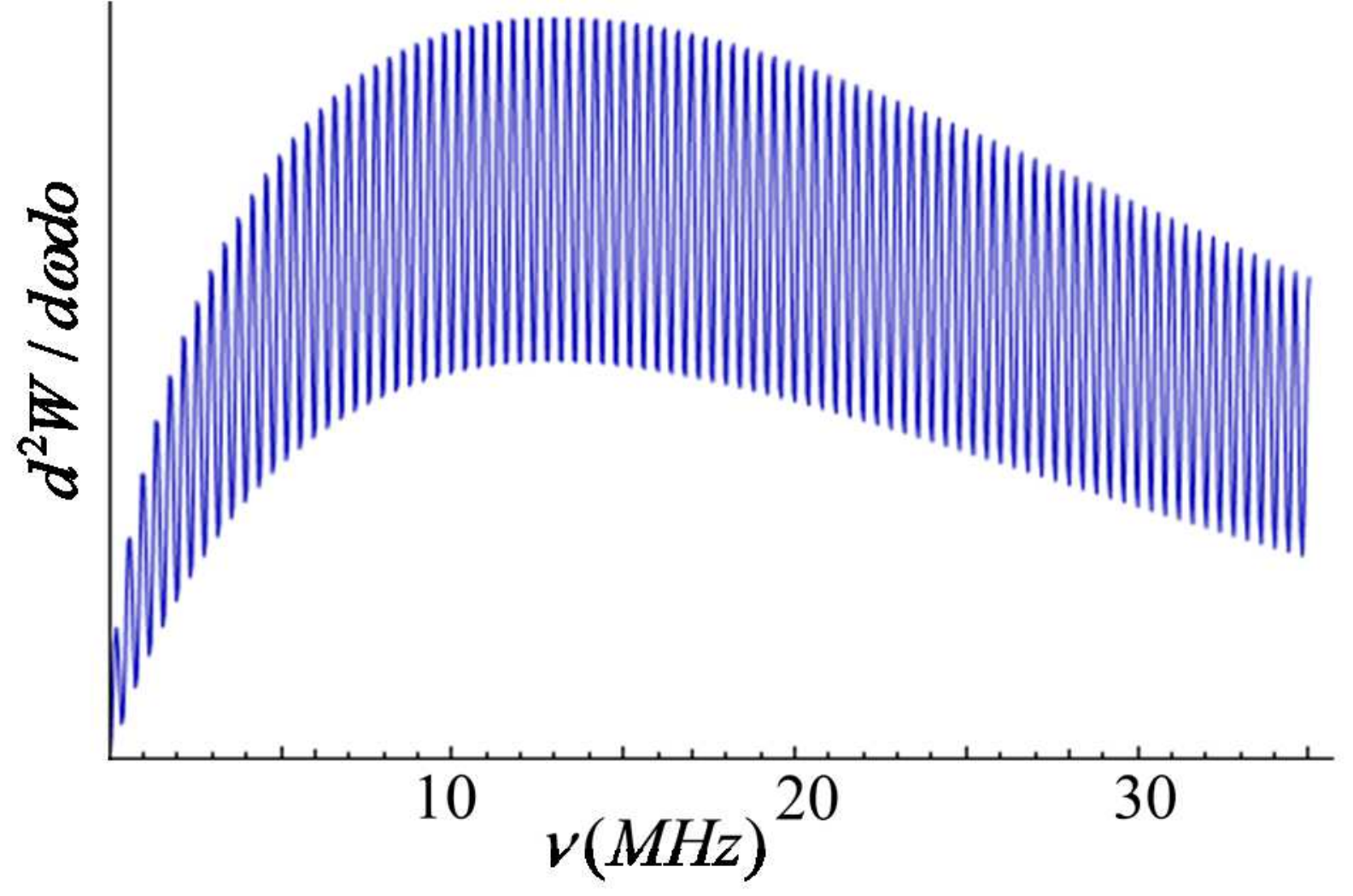}
	\end{center}
	\caption{The radiation spectral distribution (not smoothed) for $\varphi=15^\circ$. Rapid oscillations originate from the interference of the reflected curvature radiation with the transition radiation. The figure is presented for the frequencies lower than the ones we are interested in so that the oscillations were distinguishable \citep{aasp}.}
	\label{Fig6}
\end{figure}

The maximum value $\psi_{max}$ of the angle $\psi$ at which the reflected radiation still can escape from the pulsar and hit the telescope, certainly, depends on a series of parameters characterizing the magnetosphere and, probably, can not be defined from pure theoretical speculations. Fig.5, illustrating the expression (\ref{eq4}), shows that in the model under discussion the characteristic frequencies of the waves propagating in the directions corresponding to smaller values of $\psi$ are higher than $\omega(\psi_{max})\equiv\omega_{min}$. Therefore $\omega_{min}$ can be approximately considered as the minimum frequency at which the pulse of radiation from the positrons can be observed. According to (\ref{Fig6}), the value of $\omega_{min}$ is related to characteristic minimum value $\gamma_{min}$ of the positron Lorenz-factor, at which the emitted radiation begins hitting the telescope, as
\begin{equation}\label{eq6_1}
\gamma_{min}\sim(\omega_{min}R/c)^{1/3}
\end{equation} 

Let us take, for example, $\gamma_{min}\sim 200$. In this case the choice of the values $f_1=4\cdot10^3$ and $f_2=3\cdot10^7$ for the corresponding quantities in (\ref{eq3}) gives the value $\omega_{min}/(2\pi)\sim5 GHz$ for the minimum characteristic frequency at which the positron radiation pulse appears. As fig.1 shows, such result is in accordance with the observed value of the frequency at which the shifted IP appears.

It is necessary to note that fig.5 shows the frequency dependence of the averaged (over small frequency intervals) value of radiation spectral-angular density defined by (\ref{eq4}). In fact there are small-amplitude rapid oscillations of the considered quantity around the presented average value. Such oscillations are caused by interference of the reflected curvature radiation with the transition one. We decided not to present such oscillations on fig.5 since they do not play important role for our estimations. Moreover they will, probably, disappear after averaging over large number of positrons radiating simultaneously within the polar gap. Nevertheless, the example of the real positron radiation distribution is presented on fig.6. Such distribution corresponds to a larger value of the angle $\varphi$ (and, hence, smaller frequencies) for which the time of numerical calculations on the basis of the expression (\ref{eq4}) is much shorter.      

Let us also note that we have chosen the region of the linear growth of the electric field to occupy the half of the effective path. In this case the positron Lorenz-factor reaches the value $\gamma(\alpha_1)\sim10^3$ in this region.  

\section{Estimation of the flux of the reflected radiation by returning positrons}
Let us now estimate the flux of the reflected radiation produced by positrons moving along magnetic field lines towards the surface of the star above the entire area of the polar cap. The observations indicate the necessity of coherent character of radiation emission by charged particles in the pulsar magnetosphere. We will make estimation taking into account this fact and use a series of results concerning coherent radiation emission by a large number of particles discussed in detail in Appendix. For such estimation we will, as previously, consider the positron trajectories in a simplified way as arcs of circles of radius $R$ varying with the distance from the magnetic axis. 

For the estimation of CR intensity and derivation of the radiation energy spectrum at first it is enough to accept the very fact of the existence of inhomogeneous (clumped) positron flux in the gap. In this case the main contribution to radiation is made by coherently radiating volumes, which we will estimate further. The number of such volumes can be rather roughly estimated through ``division" of the total radiating volume by the coherence volume. It is also necessary to take into account the dependence of such volume on the wavelength, Lorenz-factor and the curvature of the positrons trajectories.   

The radiation spectral density at frequency $\omega$ produced by a single positron, which moves along such an arc, per unit path can be estimated as follows:
\begin{equation}\label{eq7}
W_{\omega}\sim \frac{e^2}{2\pi R c}\Big(\frac{\omega R}{c}\Big)^{1/3}. 
\end{equation}  
It directly follows from the expression for spectral density of radiative energy loss by a particle moving along a circular trajectory (see, e. g. \cite{b9}).  

The size of a volume of space which encloses amount of particles radiating coherently we will take as
\begin{equation}\label{eq8}
V_{coh}=r_{||}r_{\bot}^2\sim\gamma^2 \lambda^3/(4\pi^3).
\end{equation}  
Here the distances 
\begin{equation}\label{longdist}
r_{||}\sim\lambda/\pi 
\end{equation} 
and
\begin{equation}\label{transdist}
r_{\bot}\sim\gamma\lambda/(2\pi)
\end{equation}
define the linear dimensions of such coherently radiating volume in the direction of the particles motion and in the one perpendicular to it respectively (see Appendix). It is the square of the average number $N_{coh}\sim \kappa\cdot n_{GJ}V_{coh}$ of positrons in such volume that radiation flux produced by these particles is proportional to. Here $n_{GJ}\sim10^{11} \text{cm}^{-3}$ is the Goldreich-Julian density and $\kappa<1$ is some coefficient defining its difference from the real particle density in the returning positron flux. 

The amount of the volumes $V_{coh}$ which lie within the surface element $2\pi rdr$ of the polar cap at certain altitude is then
\begin{equation}\label{eq9}
d\textrm{\sffamily  N}_{\bot}\sim \frac{2\pi rdr}{r_{\bot}^2}= \frac{2\pi rdr}{\gamma^2 \lambda^2}(2\pi)^2.
\end{equation}   
The upper estimation for the number of such volumes falling on the polar cap surface segment of the order of $r_{\bot}^2$ per unit of time is then 
\begin{equation}\label{eq10}
\frac{d\textrm{\sffamily  N}_{||}}{dt}\sim\frac{c}{\lambda}.
\end{equation}  
The effective solid angle which encloses the considered radiation emitted by positrons with the Lorentz-factor $\gamma$ can be estimated as $\Omega_{eff}\sim\pi/\gamma^2$. This leads to the following expression for the effective area which such narrow pulse of the reflected radiation traverses on distance $d$ from the star: $S_{eff}\sim \pi d^2/\gamma^2$.

Summarizing the previous considerations we can present the expression for the flux of the reflected radiation by positrons as follows:
\begin{equation}\label{eq11}
J(\omega)\sim \int dz d\textrm{\sffamily  N}_{\bot}\frac{d\textrm{\sffamily  N}_{||}}{dt}\frac{W_{\omega}N_{coh}^2}{S_{eff}}. 
\end{equation}    
After substitution of the explicit expressions for the quantities presented here it transforms to
\begin{equation}\label{eq12}
J(\omega)\sim \frac{e^2\kappa^2 n_{GJ}^2\lambda^{3-\frac{1}{3}}}{(2\pi)^{\frac{5}{3}}\pi^3 d^2}\int\limits_{0}^{R_{PC}}  \frac{drr}{R^{\frac{2}{3}}(r)}\int dz\gamma^4(r,z),
\end{equation}  
where $z=R\alpha$ (see Fig.4) is the positron coordinate along the magnetic field line and $R_{PC}$ is the polar cap radius. 

Let us assume, as previously, the linear growth of accelerating electric field strength in the region of the positrons trajectories contributing to the observed radiation flux. In this case with the use of (\ref{eqgamma}) we can express the positron coordinate through its Lorenz-factor in the following form:
\begin{equation}\label{eq13}
dz=\sqrt{\frac{mc^2\bar{h}}{2eE(r)}}\frac{d\gamma}{\sqrt{\gamma}},  
\end{equation}   
where $\bar{h}$ is some averaged (with respect to $r$) length of the positron trajectory interval in the considered region. The quantity $E(r)$ is the accelerating electric field strength itself, which varies with the distance $r$ from the magnetic axis. We will take the explicit dependence of the electric field on this distance in the following form (cf. \cite{b14}):    
\begin{equation}\label{eq14}
E(r)=E_0(1-r^2/R_{PC}^2),
\end{equation}   
which implies vanishing of the field on the polar cap boundary. 

The dependence of the curvature radius of the magnetic field lines on $r$ in the vicinity of the surface of the star in the case of a dipole field is described by the expression:
\begin{equation}\label{eq15}
R(r)=\frac{4R_*^2}{3r},
\end{equation} 
in which $R_*\sim 10^6$ cm is the radius of the pulsar.

Substituting the expressions (\ref{eq13})-(\ref{eq15}) into (\ref{eq12}) we obtain:
\[
J(\omega)\sim \frac{\kappa^2 n_{GJ}^2e^2\lambda^{3-1/3}}{2^{5/3}\pi^{14/3}d^2R_*^{4/3}}
\sqrt{\frac{mc^2\bar{h}}{2eE_0}}\times
\]
\begin{equation}\label{eq16}
\times\int\limits_{0}^{R_{PC}} \frac{drr^{5/3}}{\sqrt{1-r^2/R_{PC}^2}}\int\limits_{\gamma_{min}(r,\omega)}^{\gamma_{max}}d\gamma \gamma^{7/2},
\end{equation} 
where the value $\gamma_{max}$, which we roughly consider as independent on $r$, is the effective value of the positron Lorenz-factor at which its radiation reflected from the surface of the star ceases to hit the telescope (Fig.7). Here $\gamma_{min}$ is some minimal value of the positron Lorenz-factor for the corresponding magnetic field line. For instance, on the outer field lines it coincides with the value defined by the expression (\ref{eq6_1}).

Here we assume that the initial angular width of the positron radiation diagram $\sim 1/\gamma_{min}$, when the particle moves at the beginning of the effective path, exceeds the value of the minimal angle $\theta_{min}$ between the magnetic axis and the direction to the telescope\footnote{As was noted previously, we assume nearly orthogonal mutual direction of the magnetic and the rotation axes.}. With the increase of the positron Lorenz-factor at lower altitudes the characteristic angle of its radiation diagram becomes less than $\theta_{min}$ (at $\gamma\sim \gamma_{max}$) and radiation ceases to be caught by the telescope. 

As the expression (\ref{eq16}) shows, the contribution to the radiation flux in our case grows with the increase of $\gamma$. Due to this fact the value of the integral in (\ref{eq16}) with respect to $\gamma$ is mostly defined by the upper limit $\gamma_{max}$ while the exact value of $\gamma_{min}$ is not very significant. Let us note that such situation is different from the well known case of synchrotron radiation of the electron component of cosmic rays with the decreasing energy spectrum. Due to such spectrum of the electron energies the integral with respect to $\gamma$ in this case is merely defined by the lower limit and the contribution to the radiation flux is associated only with $\gamma_{min}$. It leads to the well known relation between spectral indices of such cosmic radio emission sources as radio galaxies, quasars, supernova remnants etc.

\begin{figure}[h]
	\begin{center}
		\includegraphics[width = 78mm]{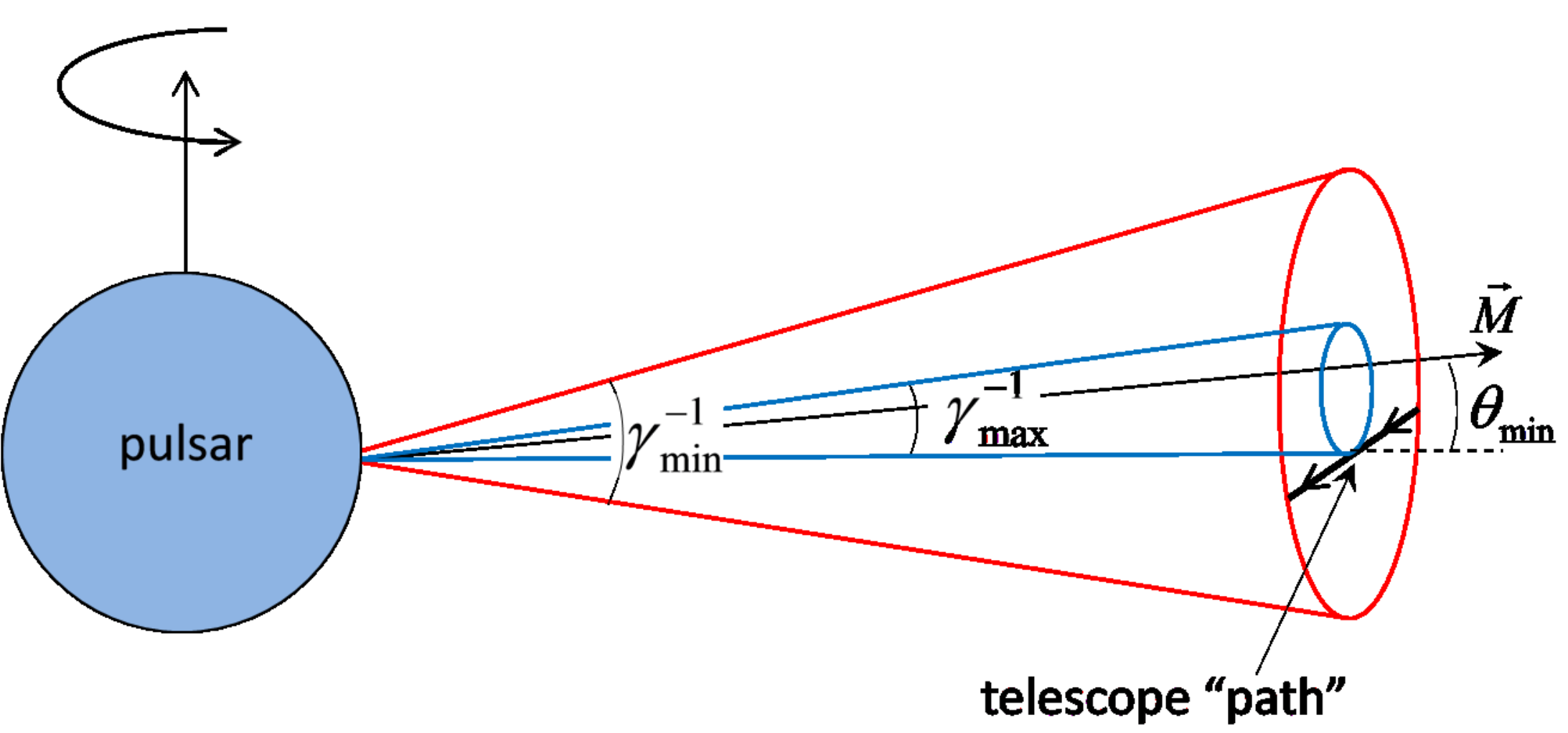}
	\end{center}
	\caption{Schematic picture of angular regions (cones) of concentration of the reflected radiation by positrons at two extreme values of the considered Lorenz-factors. At $\gamma=\gamma_{max}$ radiation ceases to hit the telescope. The magnetic axis does not coincide with the ones of the cones, among other reasons, due to its assumed inclination with respect to the surface normal}
	\label{Fig7}
\end{figure}
The final estimation of the total reflected positron radiation flux, obtained from (\ref{eq16}), is the following:
\begin{equation}\label{eq17}
J(\omega)\sim \frac{\kappa^2 n_{GJ}^2e^2\lambda^{3-1/3}}{40\pi^{14/3}d^2R_*^{4/3}}
\sqrt{\frac{mc^2\bar{h}}{2eE_0}}\gamma_{max}^{9/2}R_{PC}^{8/3},
\end{equation} 
Here we neglected the value $\gamma_{min}$ comparing to $\gamma_{max}$. 

The electric field strength $E_0$ on the magnetic axis can be estimated as: 
\[
E_0\sim a\frac{\Omega R_*}{c}B\sim10^7~\text{CGSE},
\]  
where $\Omega=2\pi/T$ is the angular frequency of the pulsar rotation, $B\sim 10^{11}$ Gs is the magnetic field strength and $a\sim10^{-1}\div10^{-2}$ is some small parameter taking into account the effect of geometrical and gravitational factors (for details see \cite{b3}).

Then for the values $R_{PC}=10^4$ cm, $\bar{h}=10^4$ cm, $d=6\cdot10^{21}$ cm (which is the distance from the Crab pulsar to the Earth (\cite{Catalogue})) of the corresponding quantities in (\ref{eq17}) this estimation reduces to the following:
\begin{equation}\label{eq18}
J(\omega)\sim \kappa^2\lambda^{3-1/3}\gamma_{max}^{9/2}\cdot10^{-40}\frac{\text{W}}{\text{Hz}\cdot\text{m}^2}.
\end{equation} 

Such estimation may, probably, be improved if apply the self-consistent calculation of the accelerating field (see ref. in \cite{b2,Muslimov}) and the returning positron flow \citep{b23}.

As the expression (\ref{eq18}) shows, the radiation flux in our model depends on a certain parameter $\gamma_{max}$ defining some average effective value of positron Lorenz-factor at which its radiation ceases to hit the telescope. Its value can be approximately estimated by the comparison of the calculated value of flux (\ref{eq18}) with the one obtained from the observations of the Crab pulsar radio emission. For the latter, for instance, the results presented in \citep{Sieber} and \citep{b27} can be applied. Namely, quite nice agreement by the order of magnitude between the results of our model and observations in the range of frequencies, in which the interpusle shift takes place, is achieved if choose $\gamma_{max}\sim 10^3$ (if take $\kappa\sim 10^{-1}$ as well).  

Expression (\ref{eq18}) also shows that in the considered here interval of wavelengths the radiation energy spectrum is defined by the term $\lambda^{3-1/3}\approx\lambda^{2.7}$, which is in quite good qualitative agreement with observation results (\cite{Sieber,b27}, for observations at lower frequencies see \cite{Ellingson}). 

Let us note that the upper estimation (\ref{eq10}) for the number of coherence volumes crossing the square $r^2_{\bot}$ per unit of time is supposed to work well for large values of the ratio $\lambda/L$. Here $L$ is some characteristic spatial size of longitudinal inhomogeneities (bunches) of the positron flow (for details see sec. I and II of Appendix devoted to partially coherent radiation). For the case of shorter waves with $\lambda/L<1$ it is necessary to change $\lambda$ to $L$ in (\ref{eq10}). In this case we have to multiply (\ref{eq18}) by $\lambda/L$ and radiation spectrum becomes proportional to $\lambda^{4-1/3}$, which even seems to be in better accordance with observations than (\ref{eq18}).

\section{Estimation of the maximum frequency at which the interpulse shift takes place}
In the framework of our model the value $\gamma_{min}$ of the positron Lorenz-factor at the beginning of the effective path defines the minimal frequency $\omega$ of its radiation which can reflect from the surface and hit the telescope. The corresponding relation between $\gamma_{min}$ and $\omega$, following from (\ref{eq6}), can be presented as:
\begin{equation}\label{eq19}
\omega\sim\frac{c}{R}\gamma_{min}^3. 
\end{equation}   
Here the radius $R$ of the positron trajectory (magnetic field line) is defined by (\ref{eq15}). Thus the value of $\gamma_{min}$ can be expressed through radiated frequency $\nu=\omega/(2\pi)$ and the distance $r$ from the magnetic field line as
\begin{equation}\label{eq22}
\gamma_{min}(\nu,r)\sim\bigg(\frac{8\pi\nu R_*^2}{3c r}\bigg)^{1/3}. 
\end{equation}
Here we see that $\gamma_{min}$ grows with the increase of $\nu$ and decrease of $r$. Therefore, at some value of $r=r_0(\nu)$ the magnitude of $\gamma_{min}$ reaches the one of $\gamma_{max}$ and the internal integral in (\ref{eq16}) tends to zero. In the region of polar cap $r<r_0$ in this case the considered radiation mechanism does not work.
According to (\ref{eq22}), the value of $r_0$ can be estimated as
\begin{equation}\label{eq23}
r_0(\nu)\sim\frac{8\pi\nu R_*^2}{3c \gamma_{max}^3}. 
\end{equation}        

\begin{figure}[h]
	\begin{center}
		\includegraphics[width = 70mm]{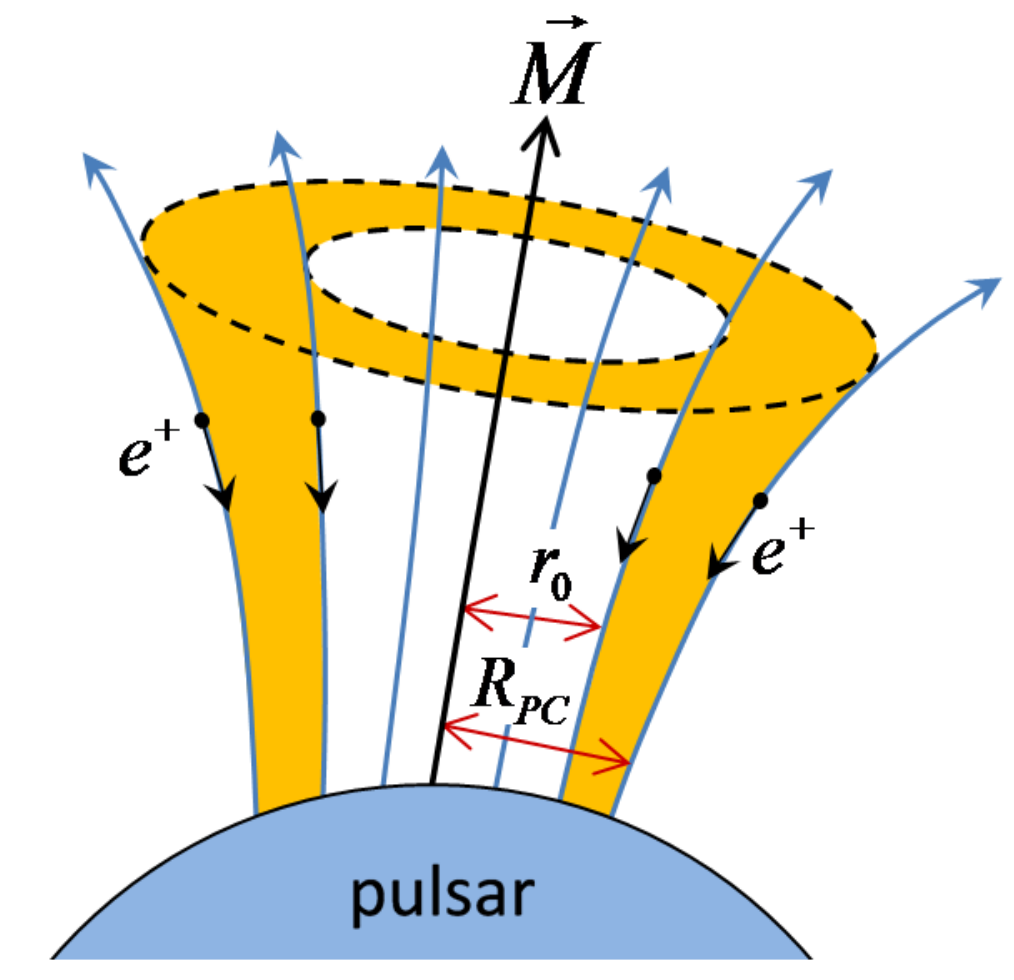}
	\end{center}
	\caption{Region of space (orange hollow cone) contributing to positron coherent radiation flux at $\nu<\nu_{max}$}
	\label{Fig8}
\end{figure}

The quantity $r_0$ is some characteristic radius of the magnetic axis vicinity, in which the moving positrons do not contribute to the observed radiation. Thus in the considered model the coherence leads to formation of something like a hollow cone (in accordance with \cite{b33}), restricted from the inner and outer sides by the field lines situated respectively on distances $r_0$ and $R_{PC}$ (in the vicinity of the star surface) from the magnetic axis (fig.\ref{Fig8}). The particles moving in the specified region make the main contribution to radiation, which corresponds to classical conceptions of pulsar radiation mechanisms.   

With the increase of the frequency $\nu$ the ``wall thickness" of the considered cone decreases ($r_0\rightarrow R_{PC}$) and the contribution to the coherent radiation is made just by more and more contracting belt in the vicinity of the outer magnetic field lines.  

The frequency $\nu=\nu_{max}$, at which $r_0$ reaches the radius $R_{PC}$ of the polar cap, is the one at which the considered radiation mechanism providing the interpulse shift disappears. With the use of (\ref{eq23}) it can be estimated as
\begin{equation}\label{eq24}
\nu_{max}\sim \frac{3c R_{PC} \gamma_{max}^3}{8\pi R_*^2}. 
\end{equation} 
Let us note that more accurate estimation of this frequency requires application of self-consistent models of the positron flow motion in the polar gap.

Taking into account the considered here effect of reduction of the region above the polar cap, which makes contribution to coherent radiation emission, with the increase of the frequency, the expression (\ref{eq17}) for the radiation flux may be slightly modified. It is associated with the fact that the integration with respect to $r$ in (\ref{eq16}) in this case should be made in the interval $r_0<r<R_{PC}$. It leads to the following expression for radiation flux:
\begin{equation}\label{eq25}
J'(\nu)\sim J\bigg[1-\frac{16}{7}x^{3/2}+\frac{9}{7}x^{8/3}\bigg],
\end{equation}
where
\[
x=\frac{r_0}{R_{PC}}=\frac{8\pi\nu R_*^2}{3c R_{PC}\gamma_{max}^3}.
\] 
and the quantity $J$ is defined by the expression (\ref{eq17}). From (\ref{eq25}) it follows that significant modification of the flux due to the considered effect takes place only at frequencies in the vicinity of $\nu_{max}$ (when $r_0\approx R_{PC}$). At lower frequencies the values of the modified expression (\ref{eq25}) and of the one, not taking into account the effect of radiation region reduction (\ref{eq17}), are nearly the same.

\section{Conclusions}
In the present work a new mechanism of radio emission in the polar gap of a pulsar is proposed. It is the radiation by positrons accelerated toward the surface of the star along curved magnetic field lines by the same electric field which accelerates electrons in the opposite direction. The total radiation by positrons in this case consists of curvature radiation reflected from the surface of the neutron star as well as of transition radiation emitted when the particles hit this surface. It is shown that the considered radiation mechanism can be applied for explanation of such unusual fact of the Crab pulsar radio emission as the shift of its interpulse at several GHz frequency. For this the assumption of the inclined magnetic field is used. The total flux of the coherent positron radiation is estimated. The obtained value (as well as the radiation spectrum) quite nicely agrees with the results of observation. In the framework of the proposed model the highest frequency at which the interpulse shift takes place is roughly estimated.

\section*{Acknowledgments}
\noindent We are grateful to E.Yu.~Bannikova, I.~Semenkina, N.F.~Shul'ga, D.P. Barsukov, O.M. Ulianov, V.V. Zakharenko and D.M. Vavriv for help and discussion.

$
$
\begin{center}
{\bf APPENDIX. On the coherence of radiation by positrons.\\
	I. General considerations}
\end{center}

The high intensity of radiation coming from the Crab pulsar (as well as from the other pulsars) claims the coherent (at least partly) emission of such radiation by a large number of particles in the pulsar magnetosphere (see, e.g. the latest review \cite{Melrose}). The possibility of coherent radiation emission by a flow of positrons moving along curved magnetic field line arises if the positron flux becomes inhomogeneous. 
  
The inhomogeneities, which lead to (partially) coherent radiation, are associated with evolution of instabilities of the bunches in magnetospheric plasma \citep{b45}. Initial inhomogeneities already appear at the electrons emission from the surface of the star or at the beginning of the positrons return motion in the lower magnetosphere (\cite{b41,Cheng,Al'ber,Asseo,Ursov}; for the analogies in the electron beams ejected from the Sun see e.g. \cite{Krasnoselskikh}). Further we apply just the very fact of the inhomogeneities existence, which makes the coherent mechanisms possible\footnote{The instabilities of the positron flow, probably, may not have time to increase up to necessary magnitude during a single passage of the gap by a positron. Nevertheless, the positrons may transfer the excitations to the electron flow moving towards them, which in its turn transfers the excitations to the positrons. Thus the positive feedback might take place, which transforms the convective instability in the system of interacting colliding beams into the absolute one, at which the excitations grow with time in each point of the beam. Such growth brings the bunches to the stationary non-linear regime, in which the maximum possible amplitudes of the inhomogeneities are established and the particle flows become clumped.}.  

%\textit{additional text with references (\citep{b45} and \citep{b35,b36} as well as \citep{Usov,Ursov})}

Further we estimate the possible spatial size of the regions responsible for coherent radiation emission by positrons in the framework of the model proposed in the present paper. Essential role in this case is played by the large transversal size of the coherently radiating area in a relativistic beam.\\

\begin{center}
	{\bf II. Longitudinal coherence}
\end{center}

Presently we will consider the situation when the density of the positron flux is inhomogeneous (the regions of the increased particle density alternate with the regions of the decreased one) in the direction of the positrons motion. Let us denote the characteristic size of the inhomogeneity as $L$. In such case it is usually expected that the radiation is emitted coherently if the wavelength exceeds the value of $L$. However, this value depends on numerous factors and, generally speaking, may significantly exceed the wavelengths $\lambda$ we are presently interested in (which are just about several centimeter). Is coherent radiation emission possible in this case? The simple calculations presented further show that partial coherence of the emitted radiation may still take place even in the case $\lambda\ll L$. By the partial coherence we mean here the case when just a certain part (probably, small) of the positrons from the region of the increased density (let us call it a bunch) are involved in coherent radiation emission.

The circular trajectory is a simplified version of the trajectory of positrons which we use for estimations in our model. Let us consider the spectral power of synchrotron (or curvature) radiation emitted in this case by a bunch of length $L$ consisting of $N$ particles. According to \cite{b37}, this quantity is defined by the following expression:
\begin{equation}\label{eqA9}
W(\omega)={\text w}(\omega)S_N,
\end{equation} 
where ${\text w}(\omega)$ is a spectral power for a single particle and $S_N$ is the so-called coherence factor. The explicit expression for it has the following form:
\[
S_N=N+N(N-1)g_{\omega},  
\]
in which the form factor
\[
g_{\omega}=\Bigg( \int\limits^{\pi}_{-\pi}d\varphi\sigma(\varphi)\cos(\omega\varphi/\omega_0) \Bigg)^2.
\]
Here $N\sigma(\varphi)$ is the density of the bunch (at the certain moment of time) as a function of the angular coordinate $\varphi$ on the circular trajectory of the bunch. The function $\sigma(\varphi)$ is supposed to be symmetric with respect to the point $\varphi=0$. 

Let us consider the simplest case of a uniform distribution of particles within the bunch. Let the bunch be situated in the region $-\varphi_0/2<\varphi<\varphi_0/2$. In this case  $\sigma(\varphi)=1/\varphi_0$ in this region while it equals zero elsewhere. In this case the coherence factor is defined by the following expression:
\begin{equation}\label{eqA7}
S_N\approx \bigg(\frac{N\lambda}{\pi L}\bigg)^2\sin^2\bigg( \frac{\pi L}{\lambda} \bigg).
\end{equation} 
Here we applied the relations $\omega_0=c/R$, $L=R\varphi_0$ and $\omega=2\pi c/\lambda$ and took into account the condition $N\gg 1$. Generally speaking, the lengths of the bunches, which the initial flux of positrons may disintegrate into, may slightly vary from one to another. If the condition $L\gg\lambda$ is fulfilled, the sine argument in (\ref{eqA7}) is a quickly oscillating function. It can be substituted by $1/2$, which is analogous to averaging of this expression over the lengths of the bunches. Finally the coherence factor takes the form:
\begin{equation}\label{eqA8}
\langle S_N\rangle\approx \bigg(\frac{N\lambda}{\sqrt{2}\pi L}\bigg)^2.
\end{equation}  
Expression (\ref{eqA8}) shows that in average the spectral power of radiation by a bunch is proportional to the squared number of particles from the bunch which are situated within the distance of about $\lambda/\sqrt{2}\pi$. In this sense the radiation by a bunch can be considered as partially coherent if the condition
\begin{equation}\label{eqA16}
\frac{1}{N}\ll \bigg(\frac{\lambda}{\sqrt{2}\pi L}\bigg)^2\ll 1
\end{equation}
is fulfilled. In this case 
\[
N\ll\langle S_N\rangle\ll N^2.
\]
Let us also note that under the considered condition ($L\gg\lambda$) the interference of radiation emitted by different bunches can be neglected.

$
$

\begin{center}
	{\bf III. Transversal coherence}
\end{center}

Previously we considered the possibility of coherent radiation emission by positrons due to inhomogeneity of their density in the direction along the positron velocity. However it is essential (see, for example, \cite{b38} and ref. in \cite{Tyutyunnik}) that in ultrarelativistic case the size of the spatial region in the direction orthogonal to the particle velocity, which is responsible for coherent radiation emission, can significantly exceed the size of the corresponding region in longitudinal direction, which, as noted previously, is of the order of radiation wavelength. Indeed, let us consider radiation emitted by two particles moving along the $z$-axis separated by distance $a$ in the direction orthogonal to it, defined by a unit vector ${\bf e}_\perp$. If we present the wave emitted by the first particle in the form $e^{i({\bf kr}-\omega t)}$, the wave emitted by the second particle in the same moment of time in the point with the same coordinate $z$ will be $e^{i({\bf k(r-a)}-\omega t)}$. Here we used a denomination ${\bf a}=a{\bf e}_\perp$. The phase difference between such waves in this case is totally attributed to the difference ${\bf a}$ of the transverse (with respect to the $z$-axis) coordinates of the particles and equals 
\begin{equation}\label{eqA10}
\delta\phi={\bf ka}=2\pi a \sin \vartheta/\lambda. 
\end{equation}
Here for simplicity we consider radiation in the plane of the particles motion. $\vartheta$ denotes the angle between the radiation direction and the $z$-axis. In ultrarelativistic case most part of the emitted radiation is concentrated in the vicinity of the value $\vartheta\sim 1/\gamma$ of this angle. From (\ref{eqA10}) we see that at such angles the condition $\delta\phi\ll 1$ for the phase difference between the waves emitted by both particles is valid if $a\ll \gamma\lambda$ or, in our notation, $a\ll r_\bot$, where $r_\bot$ is defined by the expression (\ref{transdist}). The condition $\delta\phi\ll 1$ is the condition of coherent radiation emission by the particles which for sufficiently large $\gamma$ can be fulfilled even for $a\gg \lambda$. In the directions which correspond to larger values of $\vartheta$, as follows from (\ref{eqA10}), the size of the transversal region responsible for coherent radiation emission decreases and for $\vartheta\sim 1$ it becomes $\sim \lambda$.  

If the particles move not parallel to each other and the directions of their velocities form a certain angle $\alpha$, the coherence of the particles radiation partially brakes. From the consideration presented above it follows that for qualitative estimations it can be accepted that the coherence exists if $\alpha\precsim1/\gamma$, provided the transversal distance between the particles is $a\ll r_\bot$, while for $\alpha>1/\gamma$ the radiation emission by the particles is incoherent.

The positrons moving along different magnetic field lines move not parallel to each other. As noted, this fact can modify the transversal dimensions of the region responsible for coherent radiation emission making them different from $r_\bot$. It is the case if the angular difference between the directions of the positrons motion within $r_\bot$ exceeds the value $1/\gamma$. Let us define whether such situation takes place in the considered model. We will make estimation for the region where the influence of the magnetic field lines curvature upon the radiation process is the most significant. It is the vicinity of the outer open magnetic field lines. We will also consider the case of rather high altitudes above the pulsar surface (around the beginning of the effective path) where the field lines curvature is the largest. We consider the magnetic field as close to the dipole one.  

According to the model of positron motion, accepted in the present paper, the considered region is situated on distance of the order of ten kilometers from the center of the pulsar. The positrons Lorenz-factors here we take to be of the order of $\gamma_{min}$. In this case for centimeter wavelengths (frequencies of the order of 10 GHz) the distance $r_\bot\sim\gamma\lambda$ exceeds ten meters in magnitude. 

If we choose the $z$-axis parallel to the magnetic axis and $y$-axis perpendicular to it, the equation of the line tangent to a magnetic field line in the point $(y_0,z_0)$ will have the following form (see e.g. \cite{b18}):
\begin{equation}\label{eqA11}
(3\sin\theta_0-2A^2\sin^3\theta_0/r_0^2)(y_0-y)+3\cos\theta_0(z_0-z)=0, 
\end{equation}
where $r_0=\sqrt{y_0^2+z_0^2}$ and $\theta_0=\arctan(y_0/z_0)$ are the spherical coordinates of the considered point in $yz$ plane and $A$ is the parameter of the magnetic line (its equatorial diameter). From (\ref{eqA11}) we can derive the angle $\beta$ between the magnetic field line and the $y$-axis in this point: 
\begin{equation}\label{eqA12}
\beta=\arctan(2A^2\sin^3\theta_0/r_0^2-3\sin\theta_0). 
\end{equation}
Substituting here the expression for $A$ from the equation of the magnetic field line in spherical coordinates $A=r_0/\sin^2\theta_0$, we obtain the expression for $\beta$ as a function of $r_0$ and $\theta_0$. 

As was noted, the values of $r_0$ which we consider are of the order of ten kilometers. The value of $A$ significantly exceeds this value and, for instance, for the line which divides the region of opened and closed magnetic field lines reaches the value of about several thousand kilometers. The corresponding value $\theta_0$ is $\theta_0=\arcsin\sqrt{r_0/A}\sim 0.1$. For such value of $\theta_0$ the expression (\ref{eqA12}) can be simplified and we obtain:    
\begin{equation}\label{eqA13}
\beta\approx\arctan(2/3\theta_0). 
\end{equation}
Using (\ref{eqA13}) we can define the range of $\theta_0$ on which the value of $\beta$ changes by the value $\sim 1/\gamma$:
\begin{equation}\label{eqA14}
\Delta\theta_0\sim 1/\gamma. 
\end{equation}
Such result is stipulated by the fact that at small values of $\theta_0$ which we consider the angle $\beta$ is close to the angle $\pi/2-\theta_0$ and the changes of $\beta$ and $\theta_0$ are the same by the order of magnitude. 

The distance in the direction approximately orthogonal to the direction of magnetic field lines which corresponds to the value $\Delta\theta_0$ from (\ref{eqA14}) is $\Delta r_{\theta}=r_0 \Delta\theta_0$ which is about ten meters. It is the distance in the plane $yz$ in the direction perpendicular to the magnetic lines, within which the moving positrons radiate coherently. We see that in the considered case it is of the order of $\gamma\lambda$ and we can conclude that the curvature of the magnetic field lines does not play important role in this respect. 

In order to define all three characteristic dimensions of the spatial volume in which the moving positrons radiate coherently it is still necessary to estimate its size in the direction orthogonal to $yz$ plane, which is associated with azimuthal coordinate $\varphi$ around the magnetic axis. This size can be found from the analogous condition that the angle between the velocities of positrons having different coordinates $\varphi$, which radiate coherently, should not exceed the value $1/\gamma$. Simple geometrical considerations give the condition for the maximum value $\Delta\varphi$ of the difference of the azimuthal coordinates of such positrons:
\begin{equation}\label{eqA15}
\Delta\varphi\sim \arccos\{(\cos\gamma^{-1}-\sin^2\beta)/\cos^2\beta\}. 
\end{equation}         
The distance associated with this angular difference is $\Delta r_{\varphi}=r_0\sin\theta_0\Delta\varphi$, which for the accepted values of $r_0$ and $\theta_0$ is about ten meters and also fits by the order of magnitude the value $\gamma\lambda$.

As the positrons move further towards the surface of the star and accelerate the value of $1/\gamma$ decreases. The curvature of magnetic field lines and the angles between the velocities of positrons moving along different lines decrease as well. Therefore for our estimations we assume that the relations $\Delta r_{\varphi}\sim\Delta r_{\theta}\sim r_\bot$ are approximately applicable within the whole effective path. For estimation of the total positron radiation flux we apply same assumption also for the positrons moving along the field lines situated nearer to the magnetic axis.     

Summarizing the presented considerations we conclude that the spatial volume $V_{coh}$, which is responsible for coherent radiation emission by positrons in the considered model, can be rather large. It is estimated as
\[
V_{coh}\sim \frac{\lambda}{\sqrt{2}\pi}\Delta r_{\theta}\Delta r_{\varphi}\sim \lambda^3\gamma^2/(4\pi^3)
\]
and contain quite huge number of positrons $N_{coh}$. The denominator $4\pi^3$ here comes from the more accurate application of the equation (\ref{eqA10}), while $\sqrt{2}$ is neglected in the denominator of the right-hand side. The radiation intensity will be roughly proportional to $N_{coh}^2$. The total positron radiation flux will be also proportional to the number of such volumes $V_{coh}$ which emit radiation within a unit of time. Such number is approximately estimated in sec. 5. 
	
Thus, the transversal coherence, which leads to significant dependence of the coherence volume $V_{coh}$ on the Lorenz-factor, plays a key role in the considered problem.

\end{document}